# One-Third Magnetization Plateau with a Preceding Novel Phase in Volborthite

H. Ishikawa[1], M. Yoshida[1], K. Nawa[1], M. Jeong[2], S. Krämer[2], M. Horvatić[2], C. Berthier[2], M.Takigawa[1], M. Akaki[1], A. Miyake[1], M. Tokunaga[1], K. Kindo[1], J. Yamaura[3], Y. Okamoto[1,4], and Z. Hiroi[1]

[1]Institute for Solid State Physics, The University of Tokyo, Kashiwa, Chiba, 277-8581, Japan

[2]Laboratoire National des Champs Magnétiques Intenses, LNCMI-CNRS (UPR3228),

UJF, UPS and INSA, B.P. 166, 38042 Grenoble Cedex 9, France

[3]Materials Research Center for Element Strategy, Tokyo Institute of Technology, Yokohama 226-8503, Japan

[4]Department of Applied Physics, Nagoya University, Nagoya 464-8603, Japan

We have synthesized high-quality single crystals of volborthite, a seemingly distorted kagome antiferromagnet, and carried out high-field magnetization measurements up to 74 T and $^{51}$V NMR measurements up to 30 T. An extremely wide 1/3 magnetization plateau appears above 28 T and continues over 74 T at 1.4 K, which has not been observed in previous study using polycrystalline samples. NMR spectra reveal an incommensurate order (most likely a spin-density wave order) below 22 T and a simple spin structure in the plateau phase. Moreover, a novel intermediate phase is found between 23 and 26 T, where the magnetization varies linearly with magnetic field and the NMR spectra indicate an inhomogeneous distribution of the internal magnetic field. This sequence of phases in volborthite bear a striking similarity to those of frustrated spin chains with a ferromagnetic nearest-neighbor coupling $J_1$ competing with an antiferromagnetic next-nearest-neighbor coupling $J_2$.

Frustrated quantum magnets have attracted much attention as playgrounds for realizing exotic quantum states such as a spin liquid [1-2]. There are two major sources of frustration: one is the geometry of spins that are coupled via one kind of antiferromagnetic interaction, and the other is the competition between two or more kinds of magnetic interactions. A typical example for the former is found in the spin-1/2 Heisenberg antiferromagnet on the two-dimensional kagome lattice. Theoretical studies have predicted spin liquids [3-6] or a valence bond crystal states for the ground state [7]. Experimentally, two copper minerals herbertsmithite $Zn_{1-x}Cu_{3+x}(OH)_6Cl_2$ [8-10] and vesignieite $BaCu_3V_2O_8(OH)_2$ [11-13] have been studied as candidate materials. On the other hand, a typical example of the second type of frustration is the quasi one-dimensional magnet with a ferromagnetic nearest-neighbor (NN) coupling $J_1$ competing with an antiferromagnetic next-nearest-neighbor (NNN) coupling $J_2$ along the chain. Such a $J_1$–$J_2$ chain system is expected to show a helical spin order in low magnetic fields, a spin-density wave (SDW) order in medium fields, and a spin nematic order in high fields just below the saturation of magnetization [14-17]. Particularly interesting is the spin nematic phase which corresponds to a multipolar state associated with bound magnon pairs. In a candidate compound $LiCuVO_4$, a linear field dependence of magnetization was observed before the saturation and was attributed to the spin nematic phase [18]. However, recent NMR experiments point to a possibility that it is caused by nonmagnetic defects in the Cu spin chain [19]. Thus, the presence of the spin nematic phase remains controversial.

Volborthite $Cu_3V_2O_7(OH)_2 \cdot 2H_2O$ is another copper mineral which crystallizes in a two-dimensional structure comprising distorted kagome nets consisting of two distinct sites of $Cu^{2+}$ ions, Cu1 and Cu2, separated by nonmagnetic $V_2O_7$ pillars and $H_2O$ molecules. The structure was first reported to be monoclinic with the space group $C2/m$ but later a transition into the low temperature $I2/a$ structure was found near room temperature [20-22]. A peculiar magnetic transition is observed in various experiments around 1 K [21,23-28], which is much lower than the Weiss temperature of −115 K; the low-temperature phase is called phase I. In addition, a series of magnetic field induced phase transitions accompanied by stepwise increases in magnetization are observed; phases II, III and IV appear above 4.5, 25.5 and 45 T, respectively [29-31]. At higher magnetic fields above 60 T, the magnetization tends to saturate approximately at 2/5 of the total magnetization [32] instead of 1/3 expected for isotropic or distorted kagome antiferromagnets [33-37]. Although volborthite was initially assumed to represent a distorted kagome antiferromagnet, several other spin models have been proposed later [23, 38-44]. An appropriate spin model is still unspecified and the origin of this variety of phases remains mystery. It is noted that all these features have shown up as a result of improvements in sample quality [21, 23-26], indicating that certain imperfections tend to obscure the intrinsic properties of volborthite.

In order to uncover the mystery of volborthite, we have successfully prepared high-quality, mm-size single crystals and carried out magnetization measurements up to 74 T and $^{51}$V NMR experiments up to 30 T. Two remarkably different features have been obtained compared with those in the previous study on polycrystalline samples: one is a 1/3-plateau spreading over a wide range of magnetic field above 28 T and the other is a novel phase at 23−26 T, where the magnetization shows a linear field dependence and the NMR spectra show an inhomogeneous distribution of the internal field. We argue that these phases in volborthite seem to be well described by a model, in which Cu2 spins form frustrated $J_1$–$J_2$ chains coupled via Cu1 spins in the distorted kagome net.

Growth of large single crystals of volborthite was made possible by carefully tuning preparation conditions and spending long time under a hydrothermal condition [22]. A typical crystal possesses an arrowhead shape with the surface parallel to the *ab* plane, i.e. the kagome plane, and with a twin boundary at the center of the arrowhead (Fig. 1). Single crystal X-ray diffraction measurements using synchrotron radiation source found a structural transition at 155 K from the $I2/a$ structure [21,22] into a low temperature structure with the space group of $P2_1/a$ (No. 14) (see Supplemental Material A [45]). The two structures are basically the same except that there are two kinds of crystallographically distinguished kagome layers in the $P2_1/a$ structure instead of one kind in the $I2/a$ structure. However, all the kagome layers have an identical arrangement of spin-carrying Cu $3d_{x2-y2}$ orbitals (Fig. 1), which has been uniquely determined from large differences in the Cu–O bond lengths [45].

High-field magnetization measurements were performed by the induction method using a pick-up coil in pulsed magnetic fields up to 74 T with a duration time of 4 ms generated by the non-destructive magnet [46]. High-field data was calibrated so as to reproduce the low-field data up to 7 T measured in a SQUID magnetometer (MPMS, Quantum Design). $^{51}$V-NMR experiments were carried out at LNCMI in Grenoble using a 20 MW resistive magnet. NMR spectra were collected by summing Fourier transforms of spin-echo signals at equally spaced magnetic field $B$ with a fixed resonance frequency.

Magnetization measurements were carried out on two piles of crystals grown for 30 days from the same preparation batch without a particular alignment in the plane. The measurement temperature was 1.4 K, which is above the magnetic ordering temperature of phase I (~ 1 K) but below that of phase II (~ 2 K) and phase III (above 4 K at 30 T) [31, 47]. As shown in Fig. 1, the two magnetization curves from the single crystals in magnetic fields $B$ parallel and perpendicular to the *ab* plane resemble each other, indicating a weak anisotropy, and are quite different from that of the polycrystalline samples. Each curve increases steeply around 20 T and then saturates at 30 T, followed by a small increase up to 74 T. This large increase at 20 T may correspond to the second magnetization step between phases II and III in the polycrystalline sample, though its magnitude is much enhanced. On the other hand, there is no third magnetization step at 46 T in the single crystals. It is also noted that we have observed a magnetization step at 4.5 T between phases I and II in a single crystal below 1 K (not discussed in this work) [48], which is similar to that in the polycrystalline sample [29]. Thus, differences in magnetization between the two samples are prominent only at large magnetic fields.

The nearly flat magnetization above 30 T must indicate a magnetization plateau. The small slopes may be attributed to contributions from the Van Vleck paramagnetism, which are determined by linear fitting of the curves as shown by the dashed lines in Fig. 1. The spin components at the magnetization plateaus are estimated from the intercepts of the linear fits: 0.38 and 0.36 $\mu_B$ per Cu in $B \perp$ and // *ab*, respectively, which are close to one-third of the saturation magnetization. The difference between the two values must come from the anisotropy of the Landé $g$ factor: the $g$ values of 2.28 and 2.18 in $B \perp$ and // *ab* can explain the observed magnetization values for the 1/3 plateaus, respectively. These $g$ values are typical for cuprates and consistent with the previous electron spin resonance experiments on a polycrystalline sample of volborthite, which provide axially symmetric $g$ values, $g_{//} = 2.40$ and $g_\perp = 2.04$ [49]; all the $d_{x2-y2}$ orbitals in volborthite are inclined approximately 50º from the *ab* plane.

To get information on the spin structure of the 1/3 plateau phase, $^{51}$V NMR measurements up to 30 T have been performed at 0.4 K on one single-domain piece of crystal. The magnetic field dependences of NMR spectra are plotted in Fig. 2(a) against the internal field $B_{int} = \nu_0/\gamma - B$, where $\nu_0$ is the resonance frequency and $\gamma = 11.1988$ MHz/T is the nuclear gyromagnetic ratio of $^{51}$V ($I = 7/2$). Every spectrum above 26 T appears as a single peak, indicating a relatively simple spin structure. Assuming that the couplings between a $^{51}$V nucleus and the neighboring six Cu spins are nearly equivalent, the center of gravity $M_1$ of an NMR spectrum is related to the magnetization $M$ by the relation $M = M_1/A$, where $A$ is a coupling constant $A = 0.41$ T/$\mu_B$ determined from the linear relation between the magnetic shift and the susceptibility in the paramagnetic phase. The magnetization deduced from $M_1$ at 0.4 K stays at 1/3 of the total magnetization above 28 T just as the bulk magnetization does at 1.4 K, as shown in Fig. 2(b). Note, however, that there is a specific window of fields $B = 26$-$28$ T, where the NMR spectrum appears as a single peak similar as in the plateau region, but $M_1$ as well as $M$ significantly increase toward 1/3.

Next we focus on the magnetic phases preceding the 1/3 plateau phase. Every spectrum below 22 T in Fig. 2(a), which corresponds to the field range for phase II, has a line shape of the double-horn type that is characteristic of an incommensurate helical or an SDW order. Moreover, our NMR experiments reveal that the nuclear relaxation rate $1/T_1$ shows only indiscernible anomaly near the transition temperature in phase II (see Supplemental Material B [45]). This indicates that the critical fluctuations associated with the spin order do not generate local field perpendicular to the applied field. Since the hyperfine coupling is dominantly isotropic, this means that the antiferromagnetic moments are parallel to the applied field. Therefore, realized in phase II must be a collinear SDW order, where the moments are aligned parallel to the field and their magnitudes are spatially modulated with an incommensurate periodicity, rather than a helical order that involves transverse spin polarization.

The NMR spectra in Fig. 2(a) change markedly above 22 T: the spectrum at 23.6 T takes an unusual line shape consisting of a few broad peaks, followed by a single peak above 26 T. Since the spectra between 23.6 and 25 T cannot be reproduced by a sum of those of phase II and the plateau phase, they are not due to a two-phase mixture. Therefore, this range of field should correspond to a new phase (phase N). Judging from the heavily broadened spectrum, the magnetic structure of phase N is characterized by an inhomogeneous distribution of the internal field. In addition, another interesting feature is observed in the magnetization curve at the corresponding field range. The field derivative of magnetization of Fig. 2(b) shows two kinks at 23.3 and 25.9 T and remains constant between them, that is, the magnetization is proportional to the field. Note that phase N occurs at the largest slope of magnetization below the saturation to the 1/3 plateau, as the field-derivative is maximized there.

How do we understand the appearance of this series of magnetic phases in volborthite under magnetic fields? Among the various possible spin models for volborthite, we now consider a $J_1$–$J_2$–$J'$–$J''$ model on the distorted kagome net (see Fig. 1) as the most likely. This model assumes frustrated $J_1$–$J_2$ spin chains along the $b$ axis formed by the Cu2 sites with ferromagnetic NN coupling $J_1$ and antiferromagnetic NNN coupling $J_2$, and antiferromagnetic interchain couplings $J'$ and $J''$ via the Cu1 sites. Janson and coworkers first proposed this type of model and calculated the magnitude of magnetic couplings for the high-temperature $C2/m$ structure by means of density functional theory: $J_1 = -80 \pm 10$ K (ferromagnetic), $J_2 = 35 \pm 15$ K (antiferromagnetic) and $J' = J'' = 100 \pm 60$ K [42]. Although these values have to be modified in the lowest-temperature $P2_1/a$ structure, it would be reasonable to assume that similar $J_1$–$J_2$ chains are embedded in the kagome net, because the arrangement of Cu $3d$ orbitals in the Cu2 chain is identical between the two structures. Moreover, since the Cu–O–Cu angles between Cu1 and Cu2 ions are

102° and 105°, respectively, significantly large antiferromagnetic interactions are expected for $J'$ and $J''$ [50].

In the $J_1$–$J_2$–$J'$–$J''$ model, the spin structure of the 1/3 plateau phase is most likely a ferrimagnetic state, where the Cu2 spin chains are completely polarized with the oppositely polarized intervening Cu1 spins, as schematically depicted in the inset of Fig. 2(b); the ferromagnetic $J_1$ favors uniformly aligned Cu2 spins. This ferrimagnetic spin structure is compatible with the simple NMR spectra of Fig. 2(a). As already discussed, the NMR results also indicate that the spin structure of phase II is a collinear SDW. Altogether, we find a striking similarity between the sequences of phases in volborthite and the frustrated $J_1$–$J_2$ chains: helical, SDW, nematic orders, and a 1/3 or fully saturated state occur in series with increasing magnetic field [16-17]. This suggests that phases I and N in volborthite have a helical spin and a nematic order, respectively, although we do not have direct experimental evidence yet. Note that the broadened peaks of the NMR spectra in phase N indicate the existence of non-uniform static spin moments with some disorder, which is not possible for the nematic state in the $J_1$–$J_2$ chains but could be associated with the moments on the Cu1 sites in volborthite. The detailed discussion on the NMR spectra will be given elsewhere [47]. We stress here that our results seriously call for theoretical investigation on the effects of interchain coupling between the $J_1$–$J_2$ spin chains in the distorted kagome geometry.

Finally, let us discuss what causes the very different magnetization curves in polycrystalline and single crystal samples. In Fig. 3, we compare the NMR spectrum of the single crystal in the 1/3 plateau at the field of 30 T perpendicular to the *ab*-plane [the top spectrum in Fig. 2(a)] with the spectrum of the polycrystalline sample. As discussed in ref. [31], the spectrum of the polycrystalline sample consists of two components of nearly equal intensity with different values of spin-echo decay rates $1/T_2$ (the black solid line and the blue dotted line in Fig. 3). One of them with small $1/T_2$ (solid line) shows a powder pattern for a ferromagnet or ferrimagnet due to anisotropic hyperfine couplings. We are now confident that this "slow" component is associated with the 1/3 plateau phase, because the resonance line of the single crystal for $B \perp ab$, the direction corresponding to the minimum hyperfine coupling, appears at the low field edge of the "slow" component of the polycrystalline sample (Fig. 3).

The second "fast" component of the polycrystalline NMR spectrum with large $1/T_2$ (dotted line) has a broad Gaussian-like shape, suggesting an inhomogeneous distribution of the internal field due to certain disorder. Remarkably, such a second component is almost absent in the spectrum of the single crystal, indicating much better microscopic homogeneity. Since the "fast" component has smaller values of $B_{int}$, the disordered region has smaller magnetization, consistent with the smaller magnetization of the polycrystalline sample. In fact, the centers of gravity of the "fast" and "slow" components correspond to magnetizations of 0.16 and 0.31 $\mu_B$, using the averaged $A = 0.77$ T/$\mu_B$ [26], which give a weighted average magnetization of 0.23 $\mu_B$, close to the observed value of 0.21 $\mu_B$ in the polycrystalline sample at 30 T (Fig. 1). The disorder is likely related to the arrangement of the crystal water molecules between the kagome layers, which affects the shape of Cu-O octahedra via hydrogen bonding and consequently modifies the superexchange pathways.

In summary, we successfully synthesized high-quality single crystals of volborthite and performed high-field magnetization and NMR measurements. We observe a 1/3 plateau in an unexpectedly wide field range above 28 T up to over 74 T. In addition, a novel magnetic phase called phase N is found in the field range 23−26 T, between the plateau phase and phase II (the SDW phase) at lower fields. We propose that these rich magnetic phases in volborthite come from a unique situation where frustrated $J_1$–$J_2$ spin chains are connected by intervening spins in the distorted kagome net.

We are grateful to O. Janson for helpful discussion. H. I. was supported by research fellowship of Japan Society for the Promotion of Science and its Program for Leading Graduate Schools (MERIT). Synchrotron radiation experiments were performed at BL-8A in KEK-PF (Proposal No. 2013S2-002). This work was partly supported by a Grant-in-Aid for Scientific Research (Nos. 26800176 and 25287083) and Elements Strategy Initiative from MEXT Japan and EuroMagNET II network under the EU contract No. 228043.

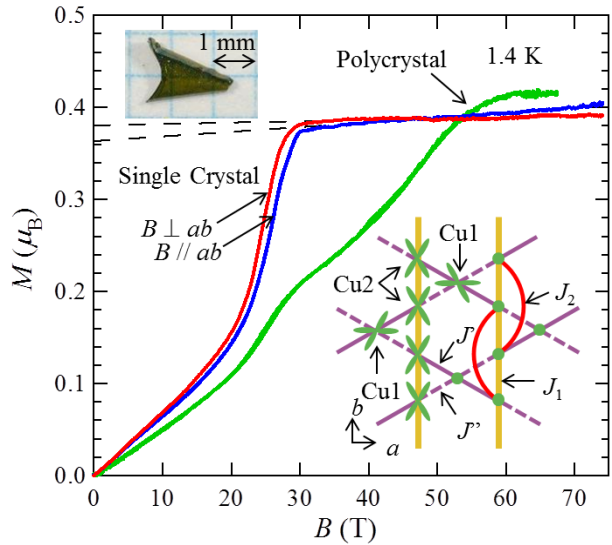

FIG. 1 (color online). Magnetization curves of volborthite measured at 1.4 K on two piles of single crystals in magnetic fields perpendicular (red) and parallel (blue) to the $ab$ plane, and on a polycrystalline sample (green, [32]). Shown also are a typical single crystal of volborthite (upper left) and the arrangement of Cu $d_{x2-y2}$ orbitals projected onto the $ab$ plane in the low-temperature $P2_1/a$ structure (lower right). $J_1$ and $J_2$ represent the NN and NNN interactions in the Cu2 spin chains, respectively. $J'$ and $J''$ represent the NN interactions between Cu1 and Cu2 spins.

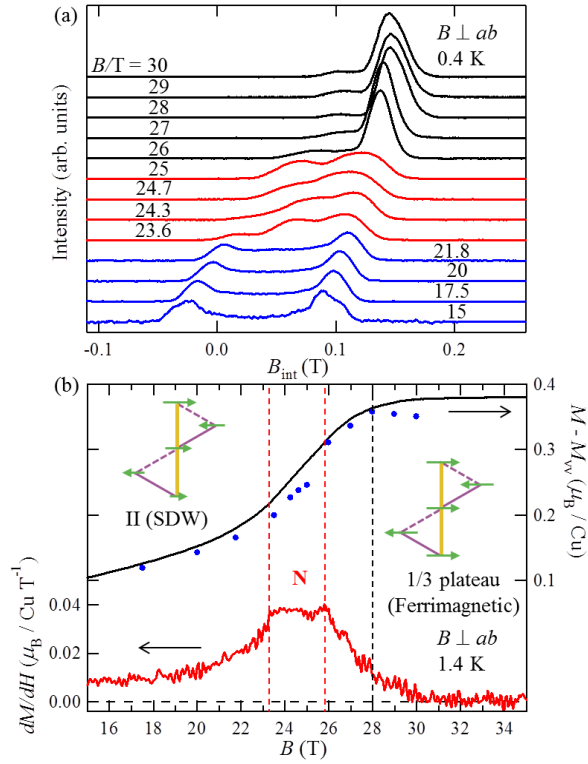

FIG. 2 (color online). (a) $^{51}$V NMR spectra measured on a single-domain piece of a crystal in magnetic fields applied perpendicular to the *ab* plane at $T = 0.4$ K. The labeled fields correspond to $B = \nu_0/\gamma$ ($B_{int} = 0$). (b) Magnetization curve of single crystals (top, black line) and its field derivative (bottom) in $B \perp ab$ at 1.4 K after the subtraction of the Van Vleck paramagnetic magnetization ($M_{VV}$). Magnetization deduced from the center of the gravity of the NMR spectra is also plotted (top, blue circles).

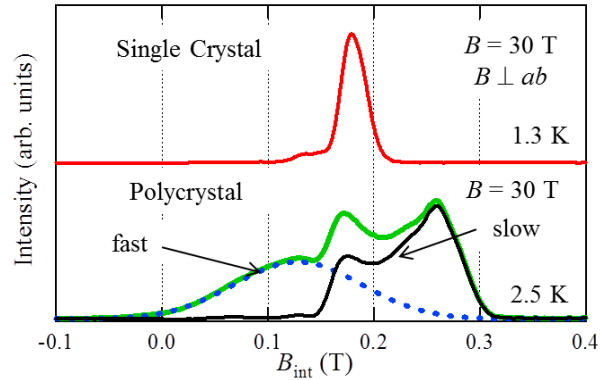

FIG. 3 (color online). NMR spectra of a single crystal at 1.3 K with the field perpendicular to the *ab* plane (top) and a polycrystalline sample (bottom, [31]). In the single crystal spectrum, the $B_{int}$ has been corrected by taking into account a demagnetization field. The powder spectrum consists of two components with different values of spin-echo decay rates $1/T_2$ as indicated by black solid and blue dotted lines (see ref. [31] for detail).

Supplemental Material

A. Structural information

Figures S1 and S2 show the two crystal structures of volborthite: the $P2_1/a$ (space group No. 14, unique axis $b$, cell choice 3) structure at low temperatures (left) and the $I2/a$ (space group No. 15, unique axis $b$, cell choice 3) structure at high temperatures (right), which were determined by single crystal X-ray diffraction experiments. The transition between them occurs at 155 K. The $P2_1/a$ phase has lattice constants of $a = 10.6489(1)$ Å, $b = 5.8415(1)$ Å, $c = 14.4100(1)$ Å, and $\beta = 95.586(1)°$ at 50 K, while the $I2/a$ phase has $a = 10.6237(3)$ Å, $b = 5.8468(1)$ Å, $c = 14.3892(7)$ Å, and $\beta = 95.3569(1)°$ at 200 K. The two structures are basically similar to each other, having Cu atoms in distorted kagome nets. One notable difference is that the former contains two kinds of kagome layers in the unit cell, while one kind in the latter structure.

Figure S2 shows the coordination environments of Cu atoms in the two kagome layers of the $P2_1/a$ structure (left) and those in the kagome layer of the $I2/a$ structure (right). Cu octahedra are heavily deformed owing to the Jahn-Teller effect. Short (1.9-2.0 Å) and long (2.3-2.5 Å) Cu-O bonds are depicted by thick solid lines and thin broken lines, respectively, in Fig. S2. Since there are always four short bonds and two long bonds in every octahedron, a spin is carried in a $d_{x2-y2}$ orbital extending to short bonds. Note that the arrangements of the $d_{x2-y2}$ orbitals in all the kagome layers are identical to each other, which is shown in the inset to Fig. 1. Moreover, we consider that magnetic interactions between Cu spins are not so different between all the kagome layers, because the Cu-Cu distances and the Cu-O-Cu bond angles take similar values. For details, look at the cif files of the two structures.

B. Nuclear spin-lattice relaxation rate $1/T_1$

Figure S3 shows the temperature dependences of the nuclear spin-lattice relaxation rate $1/T_1$ at 1 and 6 T for the polycrystalline sample previously examined [1] and at 9 T for the same single crystal used in the present experiments. In the single crystal measurements, the magnetic field was applied perpendicular to the $ab$ plane. We determined $1/T_1$ by fitting the spin-echo intensity $M_l(t)$ as a function of the time $t$ after several saturating pulses to the exponential recovery function $M_l(t) = M_{eq} - M_0 \exp(-t/T_1)$, where $M_{eq}$ is the intensity at thermal equilibrium. When this function did not fit the data well owing to inhomogeneous distribution in $1/T_1$, we used the stretched exponential function $M_l(t) = M_{eq} - M_0 \exp\{-(t/T_1)^\beta\}$ to determine the representative value of $1/T_1$. The inset of Fig. S3 shows the temperature dependences of the stretch exponent $\beta$, which indicates that an inhomogeneous distribution in $1/T_1$ occurs below ~2 K.

The $1/T_1$ at 1 T for the polycrystalline sample shows a sharp peak at 0.9 K, which indicates an enhancement in magnetic fluctuations which are associated with ordered moments *perpendicular* to the applied magnetic field. In sharp contrast, the single crystal data at 9 T in phase II shows no anomaly near the transition temperature of 2.7 K determined by the temperature dependence of the spectral width. This result indicates the absence of transverse ordered moments. A priori, one may still expect small transverse fluctuations of internal fields produced by longitudinal magnetic fluctuations through small offdiagonal components of the hyperfine coupling. However, the V sites projected onto the *ab* plane are located near the center of the hexagon formed by the Cu sites, so that small off-diagonal contributions would be cancelled out when the magnetic field is perpendicular to the *ab* plane. On the other hand, the $1/T_1$ at 6 T in phase II for the polycrystalline sample shows a small kink near 1.4 K. This must be attributed to small transverse fluctuations of the internal fields produced by longitudinal magnetic fluctuations, because the cancellation becomes impossible when the magnetic field is tilted from the normal to the *ab* plane.

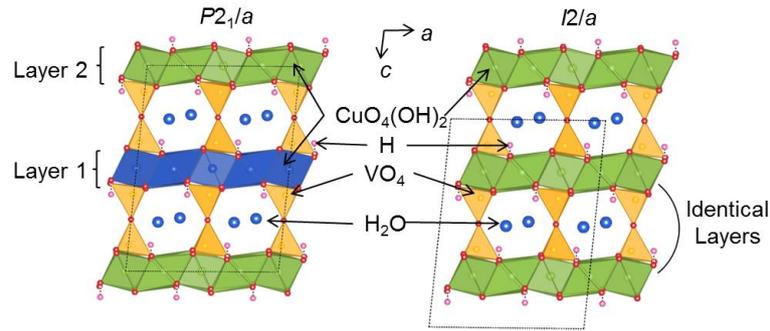

FIG. S1: Crystal Structures of the $P2_1/a$ (left) and $I2/a$ (right) structures of volborthite, which are shown by coordination polyhedra viewed along the *b* axis. The unit cells are shown by the black dotted lines.

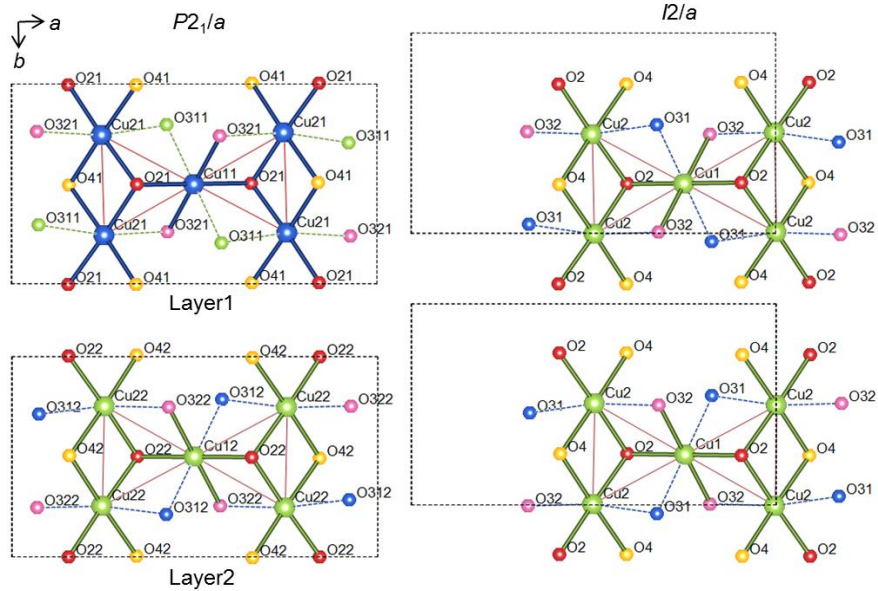

FIG. S2: Coordination environments in the kagome layers of $P2_1/a$ (left) and $I2/a$ (right) structures viewed along the *c* axis. The unit cells are shown by the black dotted lines.

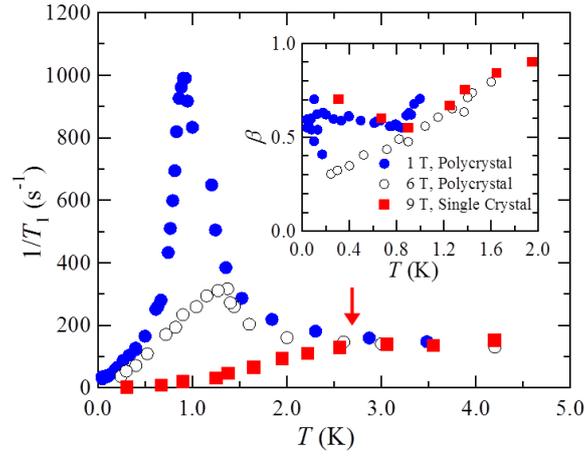

FIG. S3: Temperature dependences of $1/T_1$ at 1 T (blue dots) and 6 T (black circles) for the polycrystalline sample and at 9 T (red squares) for the single crystal. The data for the polycrystalline sample are taken from Ref. [1]. In the single crystal measurements, the magnetic field was applied perpendicular to the *ab* plane. The arrow indicates the transition temperature of 2.7 K at 9 T for the single crystal. The inset shows the temperature dependences of the stretch exponent $\beta$.